\documentclass[journal]{IEEEtran}
\usepackage{cite}

\ifCLASSINFOpdf
  \usepackage[pdftex]{graphicx}
\else
  \usepackage[dvips]{graphicx}
\fi
\usepackage{amsmath}
\usepackage{amssymb}
\DeclareMathOperator*{\argmax}{arg\max}
\usepackage{subfigure}
\usepackage{extpfeil}
\usepackage{color}
\begin{document}
%
\title{\huge{Physical Layer Authentication for Non-Coherent Massive SIMO-Enabled Industrial IoT Communications}}
%
%
%

\author{Zhifang Gu,
        He Chen,
        Pingping Xu,
        Yonghui Li, ~\IEEEmembership{Fellow,~IEEE,}
        and~Branka~Vucetic,~\IEEEmembership{Fellow,~IEEE}
\thanks{Z. Gu and P. Xu are with the National Mobile Communications Research Laboratory, Southeast University, Nanjing, China (email: {zhifang\_gu, xpp}@seu.edu.cn). H. Chen is with the Department of Information Engineering, The Chinese University of Hong Kong, Hong Kong, China (email: he.chen@ie.cuhk.edu.hk). Y. Li and B. Vucetic are with The University of Sydney, NSW 2006, Australia (email: {yonghui.li, branka.vucetic}@sydney.edu.au).}
\thanks{This work was done when Z. Gu was a visiting student at The University of Sydney under the supervision of H. Chen and Y. Li.}
}

\maketitle

\begin{abstract}
Achieving ultra-reliable, low-latency and secure communications is essential for realizing the industrial Internet of Things (IIoT). Non-coherent massive multiple-input multiple-output (MIMO) is one of promising techniques to fulfill ultra-reliable and low-latency requirements. In addition, physical layer authentication (PLA) technology is particularly suitable for secure IIoT communications thanks to its low-latency attribute. A PLA method for non-coherent massive single-input multiple-output (SIMO) IIoT communication systems is proposed in this paper. This method realizes PLA by embedding an authentication signal (tag) into a message signal, referred to as "message-based tag embedding". It is different from traditional PLA methods utilizing uniform power tags. We design the optimal tag embedding and optimize the power allocation between the message and tag signals to characterize the trade-off between the message and tag error performance. Numerical results show that the proposed message-based tag embedding PLA method is more accurate than the traditional uniform tag embedding method which has an unavoidable tag error floor close to $10\%$.
\end{abstract}


%
\IEEEpeerreviewmaketitle

\section{Introduction}
%
%
%
%
\IEEEPARstart{T}{he} industrial Internet of Things (IIoT) has  recently  attracted  tremendous  attention  from researchers  and  engineers  owing  to  its  ability  to  improve the  efficiency  and  productivity  of various industries. The IIoT combines machine-to-machine communications with automation technologies to precisely control the production process for achieving sustainable and efficient manufacturing \cite{a1}. Compared with the traditional industrial networks mainly based on wired cables, wireless communications are more suitable for the IIoT due to low maintenance expenditure, flexible deployment and long-term reliability \cite{a2}. However, ultra-reliable, low-latency and secure requirements of the IIoT represent main challenges for wireless design \cite{a3}, \cite{a4}. Wireless channels suffer from path-loss, shadowing, fading and interference, thus it is challenging  to  design  wireless  networks  to  achieve  the ultra-reliable  transmission  \cite{a5}. Moreover, the broadcast characteristic of wireless channels makes the IIoT systems more vulnerable to attacks \cite{a6}.
Non-coherent massive multiple-input multiple-output (MIMO) has recently been regarded as a promising technology to meet ultra-reliable and low-latency requirements of IIoT communications \cite{a7}, which uses multiple receive antennas to reduce the effects of fading and uncorrelated noise in wireless channels and to boost the system reliability. Besides, non-coherent massive MIMO uses energy-based modulation to achieve low latency by avoiding channel estimation and by applying fast non-coherent detection \cite{a8}, \cite{a9}. 

Two security services commonly considered in the IIoT, including integrity and authenticity, are essential in IIoT systems. Message authentication code (MAC) is a prevalent mechanism to provide these two services. Conventional systems realize message authentication by attaching a MAC to the message and this authentication process is executed above the physical layer \cite{aa1}, e.g., by the transport layer security (TLS) protocol in the transport layer and by the Wi-Fi protected access II (WPA2) protocol in the network layer. However, these conventional mechanisms may not be able to meet the stringent low-latency requirement of IIoT communications. Because short-packet transmission is one of the characteristics of the IIoT, the transmission overhead for the MAC can be large and excessive in the short packet transmission with small payload \cite{a2}, leading to relatively large delays. Two approaches were used to deal with this issue, one is lightweight security \cite{aa2} and the other is physical layer security \cite{a6}. Lightweight security mechanisms aim to save processing time by designing a cipher or protocol that only needs a small amount of computation. But the transmission overhead for security related data, e.g., MAC, still exists and increases the communication latency. On the other hand, addressing security issues from the physical layer incurs little or no transmission overhead and it represents a promising alternative in low latency communication scenarios.

Physical layer security, according to its implementation method, can be further divided into two categories. The first category is based on information-theoretic approach, which was proposed by Shannon \cite{aa3} and further developed by Wyner using the wiretap channel model \cite{aa4}. This kind of approach only guarantees data confidentiality by preventing eavesdropper from understanding the information, but other security services like data integrity and authenticity are not considered in these methods. In other words, they were designed to address passive attacks (e.g., eavesdropping) rather than active attacks (modification, masquerade or replay attack). Besides, the information-theoretic security might be unfeasible in the practical scenarios, because the channel advantages assumption does not always hold in practice \cite{aa5}. The second category is based on the signal and channel features, which aims to provide authenticity and data integrity. In IIoT scenarios, active attacks are much more harmful than passive attacks \cite{aa6}. Authentication is an effective mechanism to deal with active attacks, therefore, we focus on using the second category methods to realize physical layer authentication (PLA) in IIoT communication systems.

Generally, authentication has two kinds of meanings. One is identity authentication, which only cares if the message comes from a legitimate user. The other meaning is message authentication, which assures that the received data has not been altered by an unauthorized method and a given entity is the original source of the received data. In IIoT communication systems, data integrity and authenticity are both needed. Therefore, the authentication we intend to realize is the combination of identity authentication and message authentication. 

Existing PLA methods have two forms: passive and active \cite{a11}. Passive PLA utilizes the intrinsic features of communication systems to authenticate the transmitter, such as radio signal strength indicator, channel state information (CSI) and radio frequency fingerprints \cite{a6}. These features were thoroughly analyzed in \cite{a12} with a theoretical model and experimental validation. It is shown that the intrinsic features are not reliable in practical scenarios due to the device mobility, wireless fading channels and indistinguishable RF fingerprints. Furthermore, passive PLA only completes identity authentication but not message authentication, because it cannot detect if the received data have transmission error or have been modified by a malicious user. In contrast, active PLA refers to the methods in which the transmitter sends additional information (normally referred to as tag) for authentication at the physical layer \cite{a13}. 
Active PLA embeds a tag into the message information and does not take extra time to transmit the tag. Thanks to its potential to meet low-latency requirements, active PLA has advantages over conventional authentication methods sending message information and authentication information separately \cite{a14}. 

In addition to the advantage of low latency, the embedding method of active PLA has another important benefit: it increases the attacker's uncertainty about the tag information because the tag is ``hidden" in the message \cite{aa7}. An attacker who intends to invade into the IIoT system needs to acquire a valid secret key of the system. If the received message and tag pair cannot be completely distinguished by the attacker, then the secret key may not be obtained correctly even if the brute force analysis is used. The attacker's uncertainty about the tag improves the security of the IIoT system further. Admittedly, this kind of uncertainty also exists for the legitimate receiver. However, the goal of the legitimate receiver is different from that of the attacker, and the impacts of the uncertainty to them are different. Specifically, the legitimate receiver needs to decide whether the received message is authenticated successfully (one bit information) while the attacker tries to acquire the whole correct tag (tag-length bits information). In this context, the impacts of the tag uncertainty can be quite minor to the legitimate receiver but a challenging task to the attacker. Overall, the tag embedding design can reduce the transmission overhead and latency, but at the same time, it causes a trade-off between communication security and reliability. Through an appropriate tag embedding design, active PLA methods can secure the system while fulfilling its other communication requirements. It is worth noting that active PLA requires ``freshness mechanisms" \cite{aa8}, e.g., time stamps or sequence numbers, to resist replay attacks thereby realizing identity and message authentication. The freshness mechanisms can be easily adopted in a time-synchronized IIoT system with few overheads. Note that time stamps or sequence numbers can also be embedded into message information. In this context, almost no extra transmission time for the "freshness" is needed. The freshness mechanism part is not within the scope of this paper, and we mainly focus on the realization of active PLA.

The key issue of implementing active PLA is how to embed a tag into message information at the physical layer. Several methods dealing with this issue have been published. The tag was added as noise in \cite{a15}: different additional angle offsets to normal quadrature phase shift keying (QPSK) indicate different tag bits. 4/16 hierarchical quadrature amplitude modulation (QAM) was applied in \cite{a16} to transmit message and tag simultaneously, where a 4-QAM tag constellation is superimposed on a 4-QAM message constellation. Challenge-response PLA was introduced in \cite{a17} and the authentication information was embedded during ``challenge-and-response" process. Although these active PLA methods transmit a tag and a message at the same time, they all need to send separate pilots for channel estimation to acquire the instantaneous CSI. In \cite{a18}, a tag is embedded into the original pilot to form a new pilot, and the tag detection is completed by a correlation operation. Then the new pilot signal is used to estimate CSI for message recovery. Note that these existing active PLA methods are no longer suitable for non-coherent massive MIMO-based IIoT systems, because no estimation of the instantaneous CSI is performed in the non-coherent communication system. To our best knowledge, how to perform active PLA for non-coherent systems remains an open problem.

As the first attempt to fill this gap, in this paper we focus on designing an active PLA mechanism for non-coherent massive single-input multiple-output (SIMO) IIoT systems. Specifically, we consider an IIoT system where sensors with a single antenna transmit information to a controller with multiple antennas. As elaborated in \cite{a19}, non-negative pulse amplitude modulation (PAM) is a favorable scheme for the considered system. Based on this modulation scheme, we utilize the uniform embedding first as traditional active PLA methods do. The error performance prevents the use of the uniform embedding design. The reason is that the variance of the received signal will increase as the amplitude of the transmitted signal increases \cite{a20}. In this context, the tag embedding design does not have to be uniform as in the existing work, but needs to be optimized according to the message constellation, which is referred to as message-based tag embedding in this paper. The tag embedding design becomes a nontrivial problem as the increased tag signal power reduces the error probability of the tag while increasing the error probability of the message, leading to the error performance trade-off between the message and the tag. In addition, multiple optimization variables and complex constraints make this design more challenging. In this paper, we manage to find the asymptotic optimal tag embedding design based on given system requirements. Then we characterize the trade-off between message and tag error performance through setting different system requirements, which can provide useful insights for practical system design. The main contributions of this paper can be summarized as follows:
\begin{itemize}
	\item For the considered IIoT system and the modulation scheme, when the conventional uniform embedding scheme is used, a close-form message symbol error rate (SER) and tag SER are derived and the error floor is shown to illustrate the shortcomings and limitations of this embedding design.
	\item A new message-based embedding scheme is proposed to realize PLA for the considered system. An asymptotic optimal embedding design is given by solving a formulated optimization problem. The resulting design minimizes the tag SER while fulfilling the system requirements (i.e., power constraint, message accuracy). 
	\item When the system average power is fixed, the trade-off between message and tag error performance is demonstrated through setting different message accuracy requirements. The performance improvement of the tag SER will cause the performance of the message SER to decrease. However, numerical results show that the tag and message SER can be maintained at a quite low level (e.g., less than $10^{-6}$) when the system power and the number of receiving antennas are properly configured. 
\end{itemize}

The remaining part of the paper proceeds as follows. Section \ref{SystemModelSec} presents the system model of the proposed PLA method for non-coherent massive SIMO communications. The preliminaries of the communication system are introduced in Section \ref{SignalDesignSec}. The error performance of the uniform embedding and message-based embedding is analyzed in Section \ref{ProblemFormulationSec}. Through the error performance analysis, the optimal tag embedding design problem is solved in Section \ref{Message-basedSec}. The numerical results are provided in Section \ref{NumericalResultsSec}. Finally, Section \ref{ConclusionSec} summarizes this work.

\section{System Model}\label{SystemModelSec}
In this paper, we consider the PLA in a massive SIMO-based IIoT communication system, where a sensor with one single antenna transmits data to a controller with $N$ antennas\footnote{Note that in this system, multiple sensors can transmit data to the controller with time-division multiple access (TDMA).}. An attacker is within the range of this wireless communication system, who can receive the signals from sensors and send malicious signal to the controller. 

\begin{figure}[tbp]
	\centering
	\includegraphics[width=0.45\textwidth]{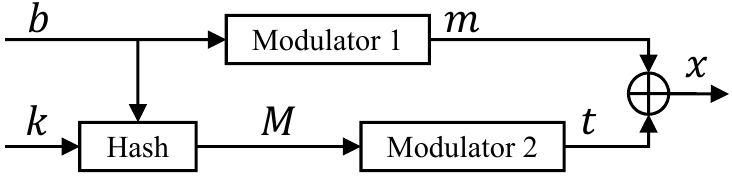}
	\caption{Diagram of the PLA at the transmitter side.}
	\label{Original Transmitter}
\end{figure}

To meet the low-latency requirement of the IIoT, we adopt non-negative PAM at the transmitter and non-coherent maximum likelihood (ML) detector at the receiver \cite{a7,a8,a9}. Since only statistics of the channel are needed in this method, the channel estimation process is not required and the authentication can be executed faster. To realize PLA, the transmitter embeds a tag signal into a message signal at the physical layer, as illustrated in Fig.~\ref{Original Transmitter}. The $l$-bit binary MAC is denoted by $M$, which is generated by a binary message sequence (denoted by $b$) and a secret key\footnote{Note that the overhead for generating and maintaining a secret key exists in both traditional methods and physical layer methods (e.g., channel-based key generation). We refer the readers to the detailed conventional key generation schemes \cite{a14} for more information.} (denoted by $k$) with a hash function. The MAC $M$ is given by
\begin{equation}
M={\rm{hash}}(b,k).
\end{equation}
The modulated signals of message $b$ and MAC $M$ are denoted by $m$ and $t$, respectively. They are referred to as the message signal and the tag signal, respectively. Since an energy-based PAM is utilized in the system, the $\oplus$ sign in Fig.~\ref{Original Transmitter} represents the sum of two signal powers. Therefore, the amplitude of the transmitted signal $x$ is then determined by
\begin{equation}
x=\sqrt{|m|^2+|t|^2},
\end{equation}
where $m\in \mathcal{M}=\{m_i|i=1, \cdots , L_m\}$ and $t\in\mathcal{T}=\{t_{i,j}|i=1, \cdots , L_m;\  j=1, \cdots, L_t\}$. $L_m$ and $L_t$ are the numbers of message signal and tag signal constellation points, respectively. The transmitted signal that involves message signal $m_i$ is denoted as $x_i$. The average power of message signal $E_m$ and the average power of tag signal $E_t$ are constrained by the total average power $E_{tot}$
\begin{equation}
E_m+E_t\le E_{tot},
\end{equation}
where $E_m=\frac{1}{L_m}\sum \limits_{i=1}^{L_m}|m_i|^2$ and $E_t=\frac{1}{L_mL_t}\sum \limits_{i=1}^{L_m}\sum \limits_{j=1}^{L_t}|t_{i,j}|^2$. Though a secret key is adopted to generate a MAC in the proposed method, it is different from traditional authentication mechanisms used at higher layers thanks to the tag embedding operation at the physical layer.

When a sensor transmits an embedded signal $x$, the received signal $\bf{y}$ at the controller in the considered massive SIMO system can be represented by
\begin{equation}
{\bf{y}}={\bf{h}}x+{\bf{n}},
\label{y}
\end{equation}
where ${\bf{h}}=[h_1, \cdots, h_N]^T$ is the SIMO channel vector and ${\bf{n}}=[n_1, \cdots, n_N]^T$ is the noise vector between the sensor and controller. We assume that $\bf{h}$ is a circularly symmetric complex Gaussian random vector. Specifically, each element of $\bf{h}$ is independently and identically distributed with a zero mean and unit variance (i.e., Rayleigh fading). Another assumption is that $\bf{n}$ is a circularly symmetric complex Gaussian random vector which is independent of $\bf{h}$. The mean vector of  $\bf{n}$  is a zero vector and its covariance matrix is $\sigma^2{\bf I}_N$ . $\sigma^2$ is assumed to be known and ${\bf I}_N$ is an $N$-dimensional identity matrix.

Two steps are required in the signal detection at the receiver side. First, following the ML detection rule for non-coherent detection, the estimated message signal $\hat{m}$ is obtained at the receiver by solving the problem
\begin{equation}
{\hat{m}}=\argmax_{m\in\mathcal{M}} {f({\bf{y}}|m)},
\label{ML1}
\end{equation}
where $f({\bf{y}}|m)$ is the probability density function (PDF) of $\bf{y}$ conditioned on $m$. Second, the estimated tag signal $\hat{t}$ is obtained based on $\hat{m}$ with the ML rule,
\begin{equation}
{\hat{t}}=\argmax_{t\in\mathcal{T}} {f({\bf{y}}|\hat{m},t)},
\end{equation}
where $f({\bf{y}}|\hat{m},t))$ is the PDF of $\bf{y}$ conditioned on $\hat{m}$ and $t$.

After the detection, we can get the estimated message sequence $b'$ and the estimated MAC $M'$ with the demodulation of $\hat{m}$ and $\hat{t}$, respectively. Then a new MAC $M_n$ is calculated with $b'$ and $k$ by the same hash function, i.e., $M_n={\rm hash}(b',k)$. The authentication process is completed by comparing $M'$ with $M_n$. The number of the same bits for binary sequences $M'$ and $M_n$ is denoted by $l'$, then the tag accuracy $\theta$ is defined as follows
\begin{equation}
\theta=\frac{l'}{l}.
\end{equation}
We use hypothesis testing to achieve authentication due to the introduced uncertainty by tag embedding. We have two hypotheses:
\begin{equation}
\begin{aligned}
   &H_0: {\bf{y}} \rm{\ does\ not\ contain\ a\ valid\ tag,} \\ &H_1: {\bf{y}} \rm{\ contains\ a\ valid\ tag.}
\end{aligned}
\end{equation}
Using $\theta$ as the test statistic, a threshold $\theta_0$ is calculated according to the system requirements. Which hypothesis should be chosen can be decided by comparing $\theta$ with $\theta_0$. If $\theta\ge\theta_0$, then $H_1$ is chosen, which represents the message can be regarded as coming from a legitimate user and has not been tampered with. Otherwise, this message will be discarded because it is not authenticated successfully. As for how to determine the threshold $\theta_0$, we adopt the Neyman-Pearson principle, which maximizes the detection rate $P(H_1|H_1)$ while the false alarm rate $P(H_1|H_0)$ is set to less than or equal to a constant $\epsilon$. Since the attacker has no knowledge of the secret key, he can only randomly generate $l-$bit MAC and embed this MAC into a message to spoof the controller. At the controller side, each received MAC bit has a probability of $\frac{1}{2}$ to be the same with the MAC bit calculated by the binary symbol of the received message and the secret key. Then the false alarm rate $P(H_1|H_0)$ can be represented by
\begin{equation}
P(H_1|H_0)\!=\!\sum_{i=\lceil{l\theta_0}\rceil}^{l}\binom{l}{i}\left(\frac{1}{2}\right)^{i}\left(\frac{1}{2}\right)^{l-{i}}\!=\!\sum_{i=\lceil{l\theta_0}\rceil}^{l}\binom{l}{i}\left(\frac{1}{2}\right)^l,
\end{equation}
where $\lceil{l\theta_0}\rceil$ is the ceiling integer of $l\theta_0$. The threshold $\theta_0$ can be obtained by solving the inequality $P(H_1|H_0)\le \epsilon$ with the following two steps. First, we find the largest integer from $0$ to $l$, which can make the inequality true. This integer is denoted by $i^*$. Second, we solve the equation $l\theta_0=i^*-1+\rho$ to obtain the threshold $\theta_0=\frac{i^*-1+\rho}{l}$, where $\rho$ is a positive infinitesimal. After obtaining $\theta_0$, we can calculate the detection rate $P(H_1|H_1)$. When a legitimate sensor transmits its signal to the controller, let $p$ denote the bit error rate of the MAC, then $P(H_1|H_1)$ can be calculated by
\begin{equation}
P(H_1|H_1)=\sum_{i=\lceil{l\theta_0}\rceil}^{l}\binom{l}{i}\left(1-p\right)^{i}\left(p\right)^{l-{i}}.
\end{equation}
When the hypothesis $H_1$ is true, $l'$ is a random variable following the binomial distribution with parameters $l$ and $1-p$, i.e., $l' \sim {\rm{B}}(l,1-p)$. According to the properties of a binomially distributed random variable, the expectation of $l'$ is $\mathbb{E}[l']=l(1-p)$. Then the expectation of the test statistic can be calculated by $\mathbb{E}[\theta]=\mathbb{E}[l']/l=1-p$. We can make a correct decision if $\theta>\theta_0$. Note that the expectation $\mathbb{E}[\theta]$ is a decreasing function of $p$. In other words, we can maximize $P(H_1|H_1)$ by minimizing $p$. In the next two sections, we will elaborate how to design tag embedding method to minimize $p$, so as to improve the security of the considered massive SIMO system.  
\section{Preliminaries and Detection Rules}\label{SignalDesignSec}
In this section, we first introduce preliminaries on non-negative PAM design for message constellation in the massive SIMO system, and then present the corresponding detection rules when a tag symbol is embedded into a message symbol.  
\subsection{Preliminaries on Message Constellation Design}\label{pre}
\begin{figure}[htbp]
	\centering
	\includegraphics[width=0.45\textwidth]{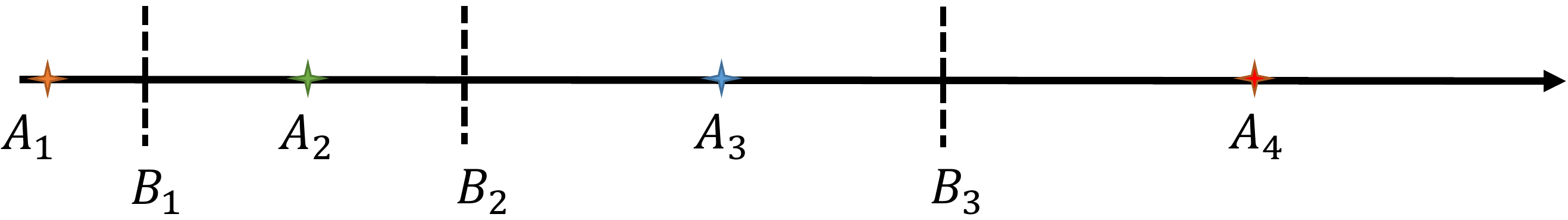}
	\caption{Message constellation design and the corresponding decision thresholds.}
	\label{Constellation Message}
\end{figure}

We now consider that only a message symbol $m\in\mathcal{M}$ is transmitted. The problem \eqref{ML1} has been solved in \cite{a9}, which showed that the ML detection problem can be solved by a quantization operation. More specifically, the quantization operation is described as \cite{a9}
\begin{equation}
\hat{m}=\begin{cases}m_{1}, & {\rm if}\ \frac{||{\bf{y}}||^2}{N}< B_1; \cr m_{i}, & {\rm if}\ B_{i-1}\le \frac{||{\bf{y}}||^2}{N}\leq B_i, \ i=2, \cdots, L_m-1; \cr m_{{L_m}}, & {\rm if}\ \frac{||{\bf{y}}||^2}{N}>  B_{L_m-1}, \end{cases}
\label{messagedetector}
\end{equation}
where $B_i$ is the optimal decision threshold\footnote{Uncoded binary sequences and hard decision are adopted in this work, an error correcting code and soft decision will be considered in the future work.} between $m_i$ and $m_{i+1}$. The threshold $B_i$ can be represented by
\begin{equation}
B_i=\frac{A_{i}A_{i+1}\ln{\frac{A_{i+1}}{A_{i}}}}{A_{i+1}-A_{i}},\quad i=1,...,L_m-1,
\label{solve_b}
\end{equation}
where $A_i=|m_i|^2+\sigma^2$. Based on this optimal decision rule, the correct detection probability of $i$-th symbol $m_i$, denoted by $P_{c,i}$, is determined by \cite{a9}
\begin{equation}
P_{c,i}=\begin{cases}G\left(\frac{NB_i}{A_{i}}\right), & {\rm if}\ i=1; \cr G\left(\frac{NB_i}{A_{i}}\right)-G\left(\frac{NB_{i-1}}{A_{i}}\right), &  {\rm if}\ i=2,\cdots,L_m-1; \cr 1-G\left(\frac{NB_{L_m-1}}{A_{i}}\right), & {\rm if}\ i=L_m,
\end{cases}
\end{equation}
where $G(z)$ is the cumulative distribution function (CDF) of a complex Chi-squared distribution variable $Z$ given by
\begin{equation}\label{CDF}
G(z)=1-e^{-z}\sum_{L=0}^{N-1}\frac{z^L}{L!},\ \ z>0.
\end{equation}
When each message symbol is selected from $\mathcal{M}$ with equal probability, the average message SER, denoted by $P_{e}$, can be calculated as follows:
\begin{equation}
\label{SERofMessage}
P_{e}=1-\frac{1}{L_m}\sum\limits_{i=1}^{L_m}P_{c,i}.
\end{equation}
By minimizing $P_{e}$ under the constraint that average message power is not greater than $E_m$, the asymptotically optimal non-negative PAM constellation design for massive SIMO systems can be represented as follows \cite{a9}:
\begin{equation}
\left\{0, \sigma^2({R}-1),\sigma^2({R}^2-1), \cdots, \sigma^2({R}^{L_m-1}-1)\right\},
\label{constellation}
\end{equation}
where  ${R}$ (${R}>1$) is obtained by solving the equation 
\begin{equation}
\sum\limits_{j=0}^{L_m-1}{R}^j=L_m\left(\frac{E_m}{\sigma^2}+1\right).
\label{solve_r}
\end{equation}
Note that the constellation design is asymptotically optimal, which means that the constellation is strictly optimal only when the signal-to-noise ratio (SNR) grows to infinity. According to the findings in \cite{a9}, this design has been shown to be close to optimal when the transmitter transmits signals with a relatively high power. In addition, when the SNR is fixed, the optimal constellation design is the same as the case with an infinite number of receiving antennas \cite{a9}. The message SNR, denoted by $\gamma_m$, is defined as ${\gamma_m}=\frac{E_m}{\sigma^2}$. Similarly, the total SNR, $\gamma_{tot}$, is defined as ${\gamma_{tot}}=\frac{E_{tot}}{\sigma^2}$. The constellation points are described from the perspective of power since there is a specific correspondence between power and amplitude in non-negative PAM. Take $L_m=4$ as an example, the message constellation points $A_i$ and the corresponding decision thresholds $B_i$ are shown in Fig.~\ref{Constellation Message}.
\subsection{Detection Rules}
Based on the asymptotically optimal message constellation design in Section \ref{pre}, we are now ready to elaborate the corresponding detection rules when a tag symbol is embedded into a message symbol. The transmitted signal $x_{i,j}$, generated by embedding $t_{i,j}$ into $m_i$, is expressed as $x_{i,j}=\sqrt{|m_i|^2+|t_{i,j}|^2}$. Due to the effect of the embedding tag signal, the message decision thresholds \eqref{solve_b} need to be updated. Since the constellation points that are close to each other are more error-prone, we use the nearest two constellation points to calculate new message decision threshold $B_i$,
\begin{equation}
B_i=\frac{A_{i,L_t}A_{i+1,1}\ln{\frac{A_{i+1,1}}{A_{i,L_t}}}}{A_{i+1,1}-A_{i,L_t}},\quad i=1,...,L_m-1,
\label{solve_b_new}
\end{equation}
where $A_{i,j}=|m_i|^2+|t_{i,j}|^2+\sigma^2$. Using \eqref{messagedetector} and \eqref{solve_b_new}, the message symbol can be estimated, which is the first step of the detection. Based on the result of estimated message symbol $m_i$, tag symbol is detected subsequently with the following rule,
\begin{equation}
\hat{t}=\begin{cases}t_{i,1}, & {\rm if}\ \frac{||{\bf{y}}||^2}{N}< C_{i,1}; \cr t_{i,j}, & {\rm if}\ C_{i,j-1}\le \frac{||{\bf{y}}||^2}{N}\leq C_{i,j}, \ j=2, \cdots, L_t-1; \cr t_{i,{L_t}}, & {\rm if}\ \frac{||{\bf{y}}||^2}{N}>  C_{i,L_t-1}, \end{cases}
\label{tagdetector}
\end{equation}
where $C_{i,j}$ is the optimal decision threshold to decide which tag symbol is embedded into message symbol $m_i$. Since \eqref{solve_b} also follows the general form of non-coherent ML decision threshold, $C_{i,j}$ can be expressed as 
\begin{equation}
\begin{aligned}
&C_{i,j}=\frac{A_{i,j}A_{i,j+1}\ln{\frac{A_{i,j+1}}{A_{i,j}}}}{A_{i,j+1}-A_{i,j}},\cr &(i=1, 2, \cdots, L_m;\ j=1,2,\cdots,L_t-1).
\label{solve_c}
\end{aligned}
\end{equation}
Note that the instantaneous CSI is not required in message detector \eqref{messagedetector} and tag detector \eqref{tagdetector}. As such, successive interference cancellation (SIC) detection widely used in existing PLA methods is no longer applicable in this method.
\section{Error Performance Analysis and Tag Embedding Methods}\label{ProblemFormulationSec}
In this section, the message SER and tag SER are analyzed first as the performance metrics of the proposed PLA method. Then two tag embedding methods are elaborated and analyzed from the perspective of the error performance.
\subsection{Error Performance Analysis}
According to the assumptions of $\bf{h}$ and $\bf{n}$ in Section \ref{SystemModelSec}, $\bf{y}$ is also a circularly symmetric complex Gaussian random vector. The mean vector of $\bf{y}$ is a zero vector, and its covariance matrix can be written as
\vspace{-1em}

\begin{small}
\begin{equation}  
\begin{split}
\mathbb{E}\left[{\bf{y}}{\bf{y}}^H\right] &=\mathbb{E}\left[\left({\bf{h}}\sqrt{|m|^2+|t|^2}+{\bf{n}}\right)\left({\bf{h}}\sqrt{|m|^2+|t|^2}+{\bf{n}}\right)^H\right] \\
&=\left(|m|^2+|t|^2+\sigma^2\right){\bf I}_N.
\end{split}
\end{equation}
\end{small}Define a new random variable $Z'$
\begin{equation}
Z'=\frac{||{\bf{y}}||^2}{|m|^2+|t|^2+\sigma^2},
\end{equation}
which follows complex Chi-squared distribution.
When $x_{1,1}$ ($m_1$ with embedding tag $t_{1,1}$) is transmitted, according to \eqref{messagedetector} and \eqref{solve_b_new}, the message signal can be correctly detected if $\frac{||{\bf{y}}||^2}{N}< B_1$. This condition is equivalent to
\begin{equation}
\frac{||{\bf{y}}||^2}{|m_1|^2+|t_{1,1}|^2+\sigma^2}< \frac{NB_1}{|m_1|^2+|t_{1,1}|^2+\sigma^2}=\frac{NB_1}{A_{1,1}}.
\end{equation}
The correct message detection probability of $x_{1,1}$ is equal to $G\left(\frac{NB_1}{A_{1,1}}\right)$. Let $P_{cm,i}$ denote the average correct message detection probability of $x_{i,j}$. Then $P_{cm,1}$ can be derived as
\begin{equation}
P_{cm,1}=\frac{1}{L_t}\sum\limits_{j=1}^{L_t}G\left(\frac{NB_1}{A_{1,j}}\right).
\end{equation}
Similarly, $P_{cm,i}$ can be given from \eqref{messagedetector} by
\vspace{-1em}

\begin{small}
\begin{equation}
P_{cm,i}\!=\!\begin{cases}\frac{1}{L_t}\!\sum\limits_{j=1}^{L_t}G\left(\frac{NB_i}{A_{i,j}}\right), & {\rm if}\ i\!=\!1; \cr\frac{1}{L_t}\!\sum\limits_{j=1}^{L_t}\!\left[G\left(\frac{NB_i}{A_{i,j}}\right)\!-\!G\left(\frac{NB_{i-1}}{A_{i,j}}\right)\!\right], &  {\rm if}\ i\!=\!2,\cdots,L_m\!-1\!; \cr \frac{1}{Lt}\!\sum\limits_{j=1}^{Lt}\left[1\!-\!G\left(\frac{NB_{L_m-1}}{A_{i,j}}\right)\right], & {\rm if}\ i\!=\!L_m.
\label{realPcm}
\end{cases}
\end{equation}
\end{small}Let $P_{em,i}$ denote the average message error rate of $x_{i,j}$, then $P_{em,i}=1-P_{cm,i}$. The average message SER $P_{em}$ can be calculated as
\begin{equation}
\label{RealmessageSER}
P_{em}=\frac{1}{L_m}\sum\limits_{i=1}^{L_m}P_{em,i}=1-\frac{1}{L_m}\sum\limits_{i=1}^{L_m}P_{cm,i}.
\end{equation}
We use the assumption that message symbols are selected from the constellation points set with equal probability. Due to the uniformity characteristic of the hash functions, different tag bits can also be considered appearing with equal probability\cite{a14}. 

We now consider the error performance of the tag signal. Suitable hash functions exhibit ``avalanche effect", i.e., the output changes significantly if the input alters slightly\cite{a14}. The authentication is extremely likely to fail when the message estimation has only one bit error, which indicates that the recalculated MAC will be meaningless if the message parts have errors. As a result, we consider the tag SER under the condition that the message symbol is detected correctly. Note that the message SER can be controlled to be low enough (e.g. less than $10^{-6}$) in a proper embedding design, where the power of tag signals is constrained according to the power of message signals. Therefore, in this paper, the tag correct rate and error rate are calculated from the perspective of conditional probability. When the embedded symbol $x_{i,j}$ is transmitted, the correct detection rate of tag $t_{i,1}$ and $t_{i,L_t}$ can be calculated by $G(NC_{i,1}/A_{i,1})$ and $1-G(NC_{i,L_t-1}/A_{i,L_t})$, respectively. When $1<j<L_t$, the correct detection rate of tag $t_{i,j}$ can be calculated by $G(NC_{i,j}/A_{i,j})-G(NC_{i,j-1}/A_{i,j})$. Therefore, the tag correct detection rate under the condition that the message symbol $m_i$ is correctly obtained, $P_{ct,i}$, can be determined by \eqref{tagdetector}, which is 
\begin{equation}
\begin{aligned}
P_{ct,i}=\frac{1}{L_t}\Bigg\{G\left(\frac{NC_{i,1}}{A_{i,1}}\right)+1-G\left(\frac{NC_{i,L_t-1}}{A_{i,Lt}}\right)\cr +\sum\limits_{j=2}^{L_t-1}\left[G\left(\frac{NC_{i,j}}{A_{i,j}}\right)\!-\!G\left(\frac{NC_{i,j-1}}{A_{i,j}}\right)\right]\Bigg\}.
\label{realPct}
\end{aligned}
\end{equation}
Therefore, the average tag SER, denoted by $P_{et}$, can be represented by
\begin{equation}
\begin{aligned}
P_{et}&=\frac{1}{L_m}\sum \limits_{i=1}^{L_m}(1-P_{ct,i})\cr
&=\frac{1}{L_mL_t}\sum \limits_{i=1}^{L_m}\sum \limits_{j=1}^{L_t-1}\left[1-G\left(\frac{NC_{i,j}}{A_{i,j}}\right)+G\left(\frac{NC_{i,j}}{A_{i,j+1}}\right)\right].
\label{Pet}
\end{aligned}
\end{equation}
\subsection{Tag Embedding Methods}
Two tag embedding methods are considered in this paper. First, we apply the uniform tag embedding method as traditional active PLA methods do. However, the error floor shown in the performance results indicates that the uniform embedding is not suitable for the IIoT system. Therefore, we propose a new method to realize PLA, which is called message-based tag embedding. The details of these two embedding methods are introduced below.
\subsubsection{Conventional Uniform Tag Embedding}\label{Uniform}
\begin{figure}[htbp]
\centering
\subfigure[Constellation pattern of uniform tag embedding.]{
\label{Constellation_uniform}
\includegraphics[width=0.46\textwidth]{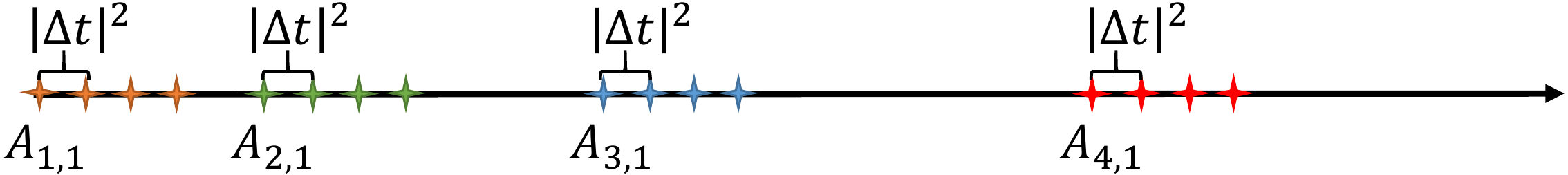}}
\subfigure[Constellation pattern of message-based tag embedding.]{
\label{Constellation_mbased}
\includegraphics[width=0.46\textwidth]{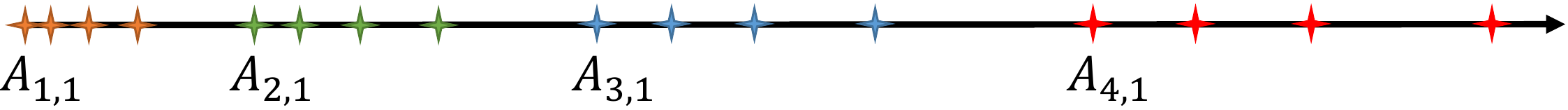}}
\caption{The constellation patterns of two tag embedding methods.}
\label{Fig.main}
\end{figure}

  Uniform tag embedding means the power interval of two adjacent tags embedded into the same message signal is a constant $|\Delta t|^2$, i.e., $|t_{i,j+1}|^2-|t_{i,j}|^2=|\Delta t|^2\ (i=1,\cdots, L_m;\ j=1,\cdots, L_t-1)$. According to the constellation design results of \eqref{constellation}, the optimal power of the first constellation point is zero. Thus, the power of the first tag symbol embedded into each message symbol is set to zero, i.e., $|t_{i,1}|^2=0\ (i=1,\cdots, L_m)$. In this context, the transmitted signal $x_{i,j}$ is represented by
\begin{equation}
    x_{i,j}=\sqrt{|m_i|^2+(j-1)|\Delta t|^2}.
\end{equation}
Take $L_m=L_t=4$ as an example, the constellation pattern of the uniform embedding method is shown in Fig.~\ref{Constellation_uniform}. 

\begin{figure}[bp]
	\centering
	\includegraphics[width=0.45\textwidth]{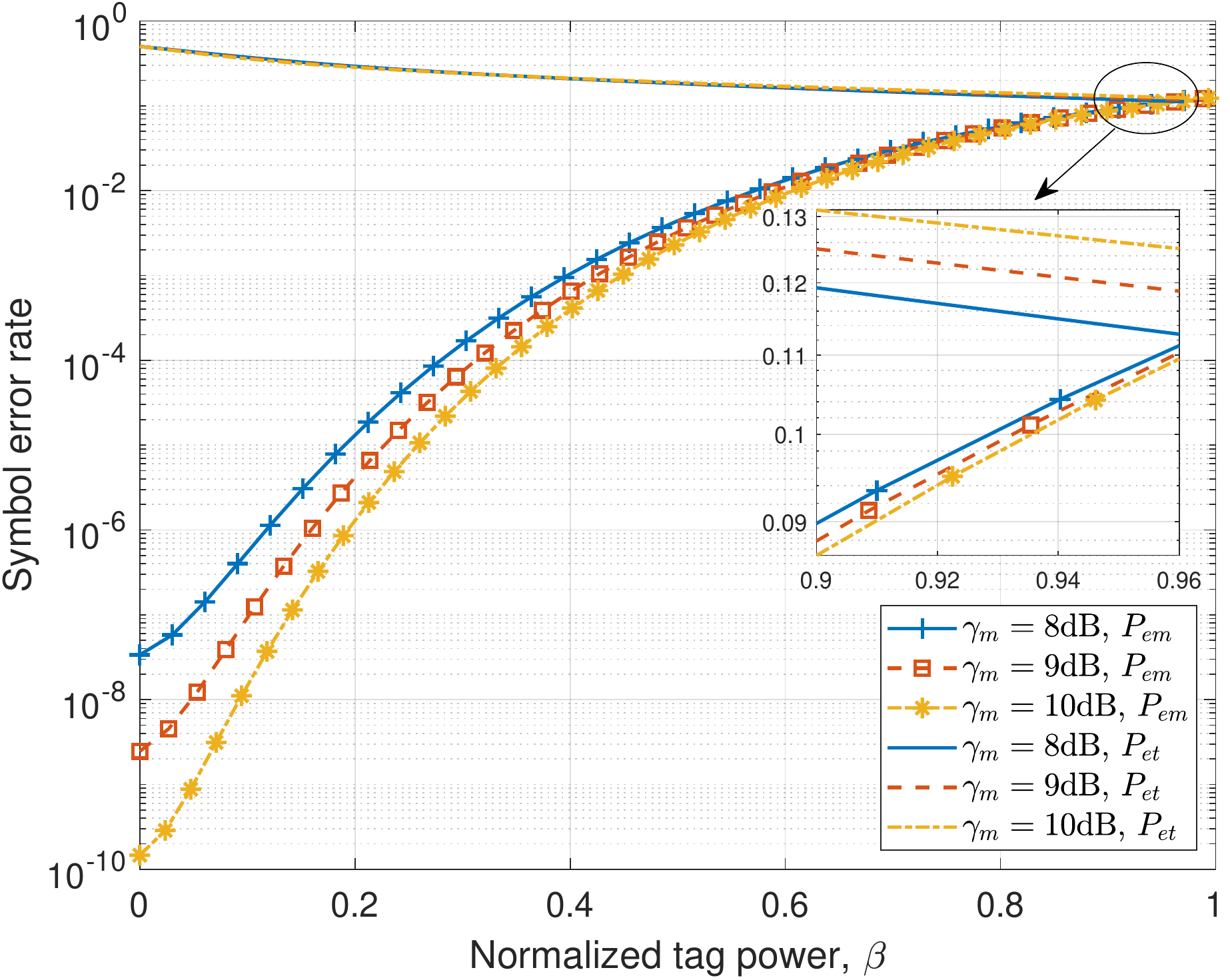}
	\caption{The error performance of the uniform tag embedding design.}
	\label{NewUniformPerformance}
\end{figure}

Using \eqref{realPcm} and $A_{i,j}=|m_i|^2+(j-1)|\Delta t|^2+\sigma^2$, we can calculate the correct message detection rate $P_{cm,i}$. Meanwhile, the correct tag detection rate $P_{ct,i}$ can be obtained by \eqref{realPct}. When $E_m$ and $L_m$ are fixed, the asymptotically optimal message constellation pattern is given by \eqref{constellation}. When we set a specific $L_t$, then $P_{em}$ and $P_{et}$ are both functions of a single variable $\Delta t$. We use a simple case ($L_m=4,\ L_t=2$, $N=128$) to examine the performance of the uniform tag embedding design. A basic rule for the uniform tag embedding is that $|\Delta t|^2$ should not exceed ${A_{2,1}-A_{1,1}}$, otherwise the power of $x_{1,2}$ will exceed the power of $x_{2,1}$ and the tag embedding will totally ruin the message signal. Let $\beta$ denote the normalized tag power, i.e., $\beta=\frac{|\Delta t|^2}{A_{2,1}-A_{1,1}}\ (\beta\le1)$. The relationship between SER and $\beta$ is shown in Fig.~\ref{NewUniformPerformance}. The message SER increases as the tag power increases. However, the tag SER only decreases slightly when the tag power increases. As shown in Fig.~\ref{NewUniformPerformance}, the tag SER has an error floor higher than $10\%$, which indicates that the uniform tag embedding is not a suitable method to realize PLA in the non-coherent massive SIMO-based IIoT system. 

\begin{figure}[tbp]
	\centering
	\includegraphics[width=0.45\textwidth]{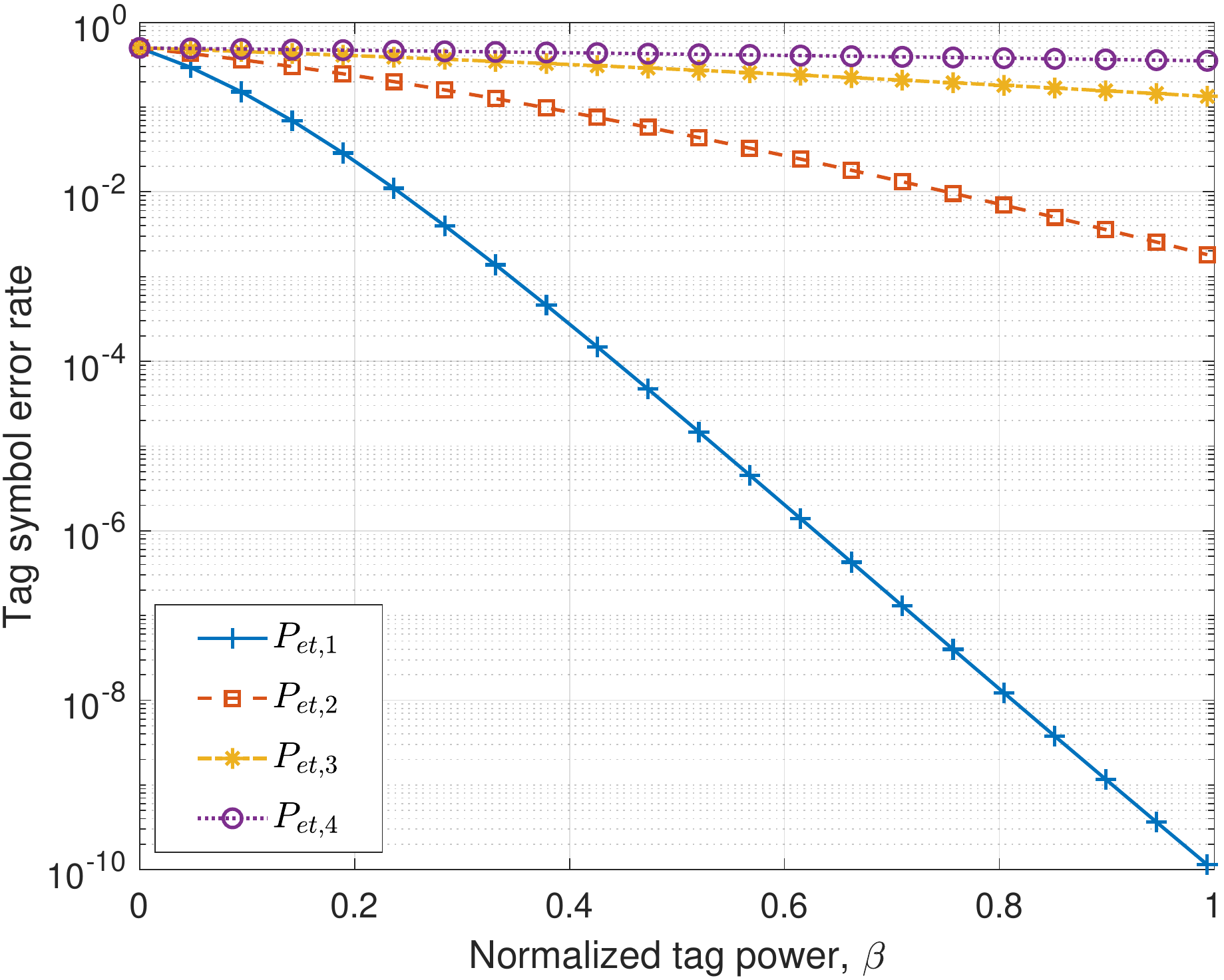}
	\caption{The SER of tags embedded into different message symbols.}
	\label{UniformPerformance}
\end{figure}

\begin{figure}[bp]
	\centering
	\includegraphics[width=0.45\textwidth]{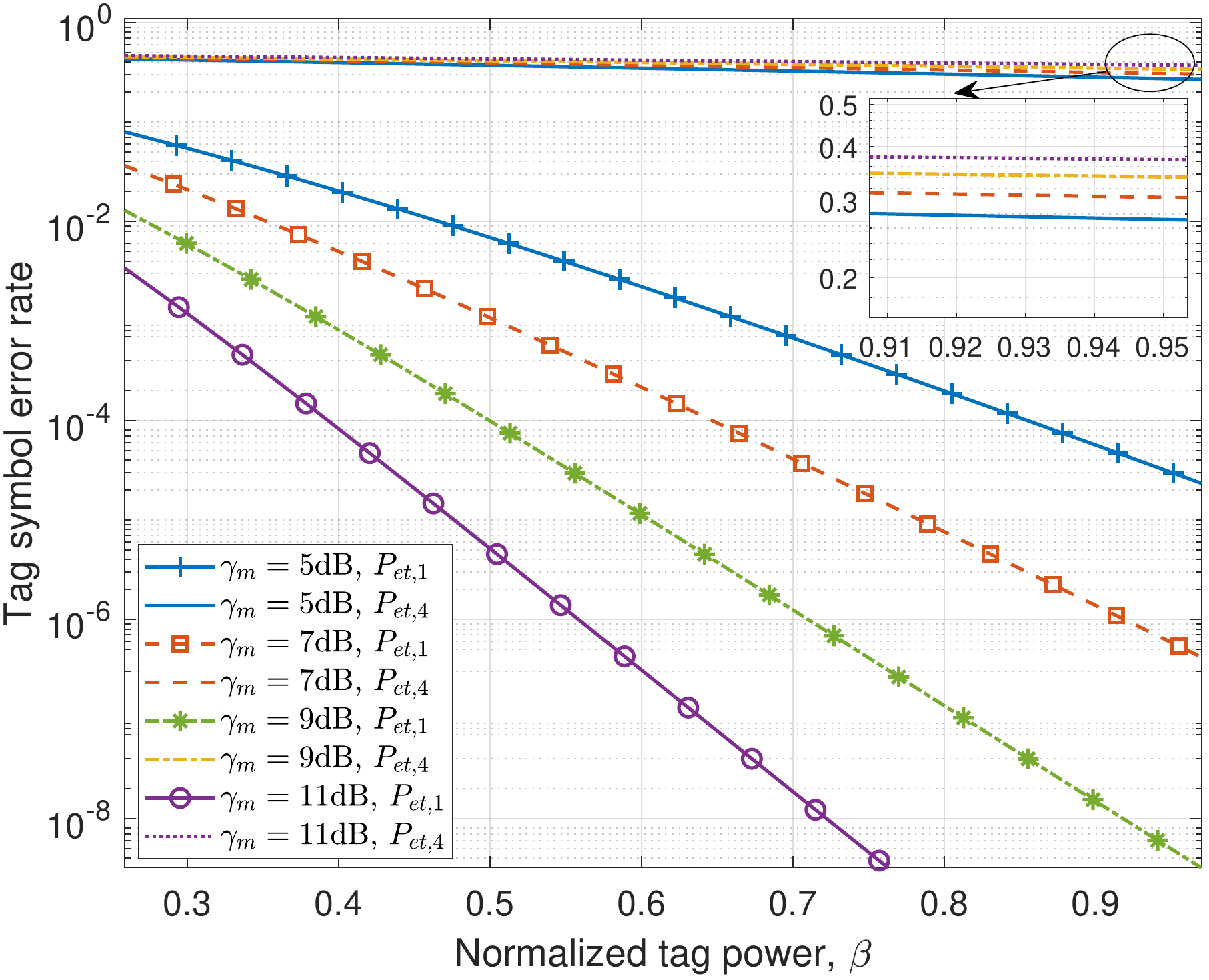}
	\caption{The SER of tags embedded into the first and the fourth symbol.}
	\label{SecondRoundPet}
\end{figure}

To analyze the cause of the error floor, we examine the SER of tags embedded into different message symbols, i.e., $P_{et,i},\ (i=1,\cdots,L_m)$. Under the same case (i.e., $L_m=4,\ L_t=2$,\ $N=128$), the trend of different tag SER  are shown in Fig.~\ref{UniformPerformance} when $\gamma_m=10$dB. In Fig.~\ref{UniformPerformance}, as the embedded tag power increases, the SER of tags embedded into different message symbols all decrease. However, the embedded tag has different degrees of impacts on each message symbol. Specifically, tags embedded in high power message symbols have much higher error probabilities than in low power message symbols, i.e., $P_{et,4}>P_{et,3}>P_{et,2}>P_{et,1}$. The reason is that the message constellation points in the considered system contain the relationship of geometric series as shown in Fig.~\ref{Constellation Message}. The embedded tag power is constrained by the lowest two power message symbols ($m_1$ and $m_2$), i.e., $|\Delta t|^2\le A_{2,1}-A_{1,1}$. In this context, the tag power is relatively low and not suitable for a high power message symbol ($m_3$ or $m_4$), because the variance of the received signal power increases as the power of the transmitted signal increases\cite{a20}. The error floor is caused by high tag error rate of $P_{et,3}$ and $P_{et,4}$. To support this analysis, we examine the value of $P_{et,1}$ and $P_{et,4}$ for different values of $\gamma_m$. From the results shown in Fig.~\ref{SecondRoundPet}, we can observe that the values of $P_{et,4}$ are above 0.2 for different $\gamma_m$. Though the value of $P_{et,1}$ decreases sharply as $\gamma_m$ increases, the error floor in Fig.~\ref{NewUniformPerformance} cannot be avoided due to a large value of $P_{et,4}$. Therefore, the uniform tag embedding design cannot achieve a good performance in the considered system. This finding motivates us to propose another tag embedding design which is called message-based tag embedding and elaborated in the subsequent subsection.  

Note that although the uniform tag embedding method was applied in some existing works \cite{a15} \cite{a16}, its application in non-coherent communication systems has not been studied prior to the work of this paper. A tag SER error floor of this method is also a new research finding.   

\subsubsection{Message-based Tag Embedding}\label{Message-basedSubSec}
Message-based tag means the power of the tag signal is dependent on the message which the tag is embedded into, in other words, the power of $t_{i,j}$ should be designed properly according to the power of $m_i$. It is reasonable to propose such tag embedding design, because the impacts of the tag power on different message symbols are different as shown in Fig.~\ref{UniformPerformance}. To be specific, the value of $t_{i,j}$ should become larger as $i$ increases due to the increasing variance of the received signal power. Lower power tags may be suitable for low power message symbol $m_1$, but they cannot perform well for large power message symbol $m_4$. On the other hand, relatively large power tags may have severe interference to low power message symbols as the decision region is narrow between low power message symbols. Therefore, message-based tag embedding is necessary in the considered massive SIMO system.

In tag set $\mathcal{T}=\{t_{i,j}|i=1, \cdots , L_m; j=1, \cdots, L_t\}$, we let $|t_{i,1}|^2=0\ (i=1,\cdots, L_m)$ as discussed in Subsection \ref{Uniform}. Then $\mathcal{T}$ has $L_m\cdot (L_t-1)$ non-zero elements, which indicates there are $L_m\cdot (L_t-1)$ variables that need to be considered in the tag embedding design. To reduce the scale of this design problem, we make an assumption that the power of embedded signal to the same message $m_i$ also consists of the geometrical sequence as expressed in \eqref{constellation}, i.e., $A_{i,j+1}=A_{i,j}r_i\ (i=1, \cdots , L_m;\ j=1, \cdots, L_t-1)$. $r_i$ is called tag embedding coefficient. This assumption is reasonable because the geometrical sequence relationship between constellation points is asymptotic optimal in the considered non-coherent PAM \cite{a9}. The value of $r_i$ ranges from 1 to $R^{1/(L_t-1)}$, i.e., $1<r_i<R^{1/(L_t-1)}$. This is because the power of a message signal plus the power of its corresponding tag signal should not exceed the power of next message signal in the constellation. In this context, the number of variables reduces from $L_m\cdot (L_t-1)$ to $L_m$. The tag average SER $P_{et}$ in \eqref{Pet} can be further written as 
\begin{equation}
\begin{aligned}
    P_{et}&=\frac{L_t-1}{L_mL_t}\sum \limits_{i=1}^{L_m}\left[1-G\left(\frac{Nr_i\ln{r_i}}{r_i-1}\right)+G\left(\frac{N\ln{r_i}}{r_i-1}\right)\right]\cr&=\frac{L_t-1}{L_mL_t}\sum \limits_{i=1}^{L_m}\left\{1-G\left[Nv(r_i)\right]+G\left[Nu(r_i)\right]\right\},
\label{FinalFormPetMessageBased}
\end{aligned}
\end{equation}
where $v(r_i)=\frac{r_i\ln{r_i}}{r_i-1}$ and $u(r_i)=\frac{\ln{r_i}}{r_i-1}$.
If we take $L_m=L_t=4$ as an example, the constellation pattern of the message-based embedding method is shown in Fig.~\ref{Constellation_mbased}. The core issue of this embedding design is how to determine $r_i\ (i=1, \cdots , L_m)$, which will be elaborated in Section \ref{Message-basedSec}.
\section{Message-based Tag Embedding Design}\label{Message-basedSec}
In this section, we elaborate how to design the constellation pattern of the message-based tag embedding method while satisfying key system requirements (e.g. communication reliability and power constraint). Then we analyze the trade-off relationship between message SER and tag SER, which expresses the impacts on authentication to system communication reliability.
The purpose of this section is to design an optimal embedding scheme to minimize tag SER when the total average power is constrained. Meanwhile, the system reliability should meet a certain requirement, i.e., $P_{em}<\delta$, where $\delta$ is the message SER requirement threshold. This optimization problem is formulated as follows
\begin{subequations}\label{generaloptimal}
\begin{align}
&\min \limits_{\{ A_{i,j}\}_{i=1,\cdots,L_m}^{j=1,\cdots,L_t}} \quad P_{et}\\ \label{general1}
&{\rm {s.t.}}\quad
E_t+E_m\leq E_{tot} , \\ \label{general2}
&\ \ \ \ \ \ \ P_{em} \le \delta.
\end{align}
\end{subequations}

This problem is non-trivial to solve because of multiple optimization variables and complex structure of the constraint function. To simplify this problem, we use two steps to solve it. First, using the assumption ($A_{i,j+1}=A_{i,j}r_i$) in Subsection \ref{Message-basedSubSec}, we find the optimal tag embedding scheme when the message constellation is fixed (i.e., $E_m$ is given). Second, we search for the optimal allocated power of message $E_m$ that can minimize the tag SER. 

We now consider the first step. When $E_m$ is fixed, the message SER $P_{em}$ and the tag SER $P_{et}$ are given by \eqref{RealmessageSER} and \eqref{FinalFormPetMessageBased}, respectively. Note that $P_{em}$ is complicated to compute since it contains multiple variables ($A_{i,j}$, $i=1,\cdots,L_m,\ j=1,\cdots,L_t$). To reduce the complexity of constraint \eqref{general2}, an upper bound of message SER, $P_{em}^u$, can be derived to replace $P_{em}$ in \eqref{general2}.  The upper bound can be derived as follows
\begin{equation}
P_{em}^u=\frac{1}{L_m}\sum \limits_{i=1}^{L_m-1}\left\{1-G\left[Ng(r_i)\right]+G\left[Nh(r_i)\right]\right\},
\label{Pemupperbound}
\end{equation}
where $g(r_i)=\frac{{R}\ln{\frac{{R}}{{r_i}^{L_t-1}}}}{{R}-{r_i}^{L_t-1}}$, $h(r_i)=\frac{{r_i}^{L_t-1}\ln{\frac{{R}}{{r_i}^{L_t-1}}}}{{R}-{r_i}^{L_t-1}}$.The proof is provided in Appendix \ref{ProofUpperBoundSec}.

Using variable substitution $r_i=e^{k_i}\ (0<k_i<\frac{\ln{R}}{L_t-1})$ and the message SER upper bound in \eqref{Pemupperbound}, the optimization problem \eqref{generaloptimal} can be reformulated as
\begin{subequations}\label{convex}
\begin{align}\label{convex1}
&\min \limits_{\{ k_i\}_{i=1}^{L_m}} \quad \frac{1}{2L_m}\sum \limits_{i=1}^{L_m}\left\{1+G\left[Nu(e^{k_i})\right]-G\left[Nv(e^{k_i})\right]\right\}\\ \label{convex2}
&{\rm {s.t.}}\quad
\frac{1}{2L_m}\sum \limits_{i=1}^{L_m}A_{i,1}(e^{k_i}-1)\leq E_{tot}-E_m, \\ \label{convex3}
&\ \ \ \ \ \ \ \frac{1}{L_m}\sum \limits_{i=1}^{L_m-1}\left\{1-G\left[Ng(e^{k_i})\right]+G\left[Nh(e^{k_i})\right]\right\} \le \delta,\\
&\ \ \ \ \ \ \ \ 0<k_i<\frac{\ln{R}}{L_t-1}.
\end{align}
\end{subequations}
We can show that \eqref{convex} is a convex optimization problem and the proof is provided in Appendix \ref{ProofConvexSec}. Therefore it can be efficiently solved by interior-point method and the optimal tag embedding scheme can be determined when $E_m$ is given.
    
In the second step, we consider the situation when $E_m$ is a variable, different $E_m$ results in different values of $P_{et}$. In this case, the optimized result of \eqref{convex} is a function of $E_m$, which is denoted by $H(E_m)$. Note that when $H(E_m)$ achieves its minimal value, the inequality \eqref{general1} will become an equation, i.e., $E_m+E_t= E_{tot}$. The reason is that if $E_{tot}$ still has a surplus, $P_{em}$ and $P_{et}$ can be further reduced by increasing message and tag power at the same time. Therefore, the second step of this optimization problem can be described as finding the optimal power allocation between message signal and tag signal when the constraints are still satisfied. Define the power allocation factor $\alpha$,
\begin{equation}
\alpha=\frac{E_m}{E_{tot}},\ \alpha_0\leq \alpha \leq1,
\end{equation}
where $\alpha_0$ is the minimum factor that makes $E_m$ satisfy the constraint $P_{em}^u \le \delta$.
The function $H(E_m)$ can be represented by $H(\alpha)$, which indicates that different $\alpha$ corresponds to different minimized tag SER. Therefore, the power allocation problem can be formulated as 
%
\begin{subequations}
\begin{align}
\label{power}
&\min \limits_{\alpha_0 \leq \alpha\leq 1}\quad H(\alpha)\\ \label{power1}
&{\rm {s.t.}}\quad
E_t= (1-\alpha)E_{tot} , \\ \label{power2}
&\ \ \ \ \ \ \ P_{e,m}^u \le \delta.
\end{align}
\end{subequations}
This is a problem of single variable with a limited range, which can be efficiently solved by one-dimensional search.

When we set different values of message SER requirement threshold $\delta$ for the considered system, different minimized tag SER can be obtained by solving the optimization problem \eqref{generaloptimal}. Therefore, we can characterize the trade-off between the message and tag error performance through changing the values of the message SER requirement thresholds. 
\section{Numerical Results}\label{NumericalResultsSec}
\begin{figure}[htbp]
	\centering
	\includegraphics[width=0.45\textwidth]{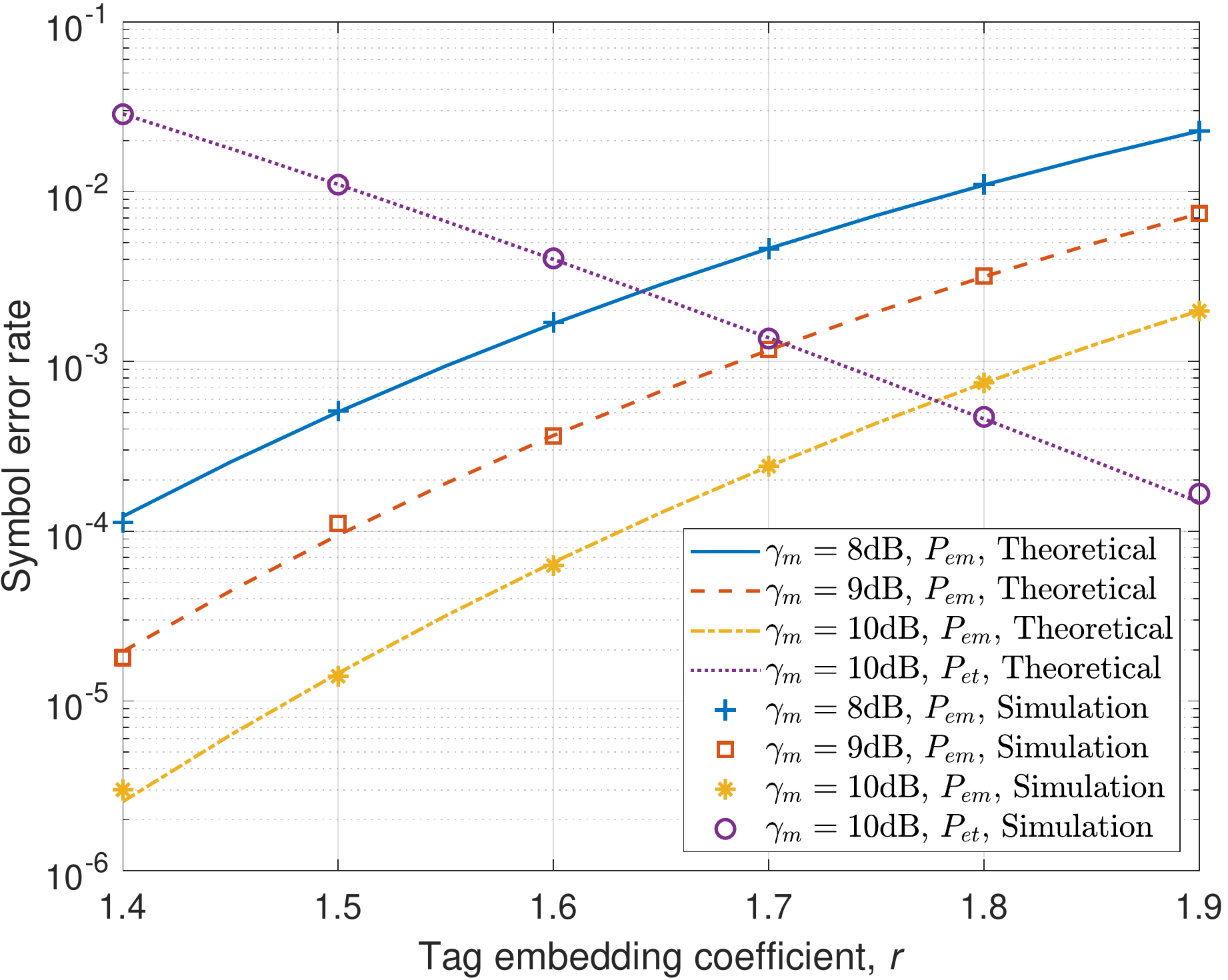}
	\caption{Error rate performance of the message-based tag embedding design.}
	\label{Message-basedPerformance}
\end{figure}

We carried out computer simulations to verify the analysis and effectiveness of our design. 
To show the accuracy of the message and tag error performance analysis of the message-based tag embedding method, the theoretical and simulation results of $P_{em}$ and $P_{et}$ are demonstrated in Fig.~\ref{Message-basedPerformance}. The simulation parameters of Fig.~\ref{Message-basedPerformance} are set the same as that in Fig.~\ref{NewUniformPerformance} to compare the proposed PLA method with the uniform tag embedding, i.e., $N=128,\ L_m=4,\ L_t=2$. We can see from Fig.~\ref{Message-basedPerformance} that the simulation results match well with the theoretical counterparts, which verifies the accuracy of the theoretical analysis. Therefore, we use theoretical results in the following figures to show other properties. As shown in Fig.~\ref{Message-basedPerformance}, the message SER $P_{em}$ increases as the tag embedding coefficient $r$ increases. Meanwhile, the tag SER $P_{et}$ decreases sharply, which is a contrast to the tag error performance of the uniform embedding method in Fig.~\ref{NewUniformPerformance}. To be specific, the tag SER of message-based embedding can decrease continuously while the tag SER of uniform embedding has an error floor close to $0.1$. Note that $P_{et}$ is independent of $\gamma_m$ as expressed in \eqref{FinalFormPetMessageBased}, and we only show the theoretical and simulation results of $P_{et}$ when $\gamma_m=10$dB. Another key observation from Fig.~\ref{Message-basedPerformance} is that the message SER $P_{em}$ decreases significantly when $\gamma_m$ increases, because the distance between adjacent constellation points gets farther and the communication error probability reduces accordingly. 

\begin{figure}[tbp]
	\centering
	\includegraphics[width=0.45\textwidth]{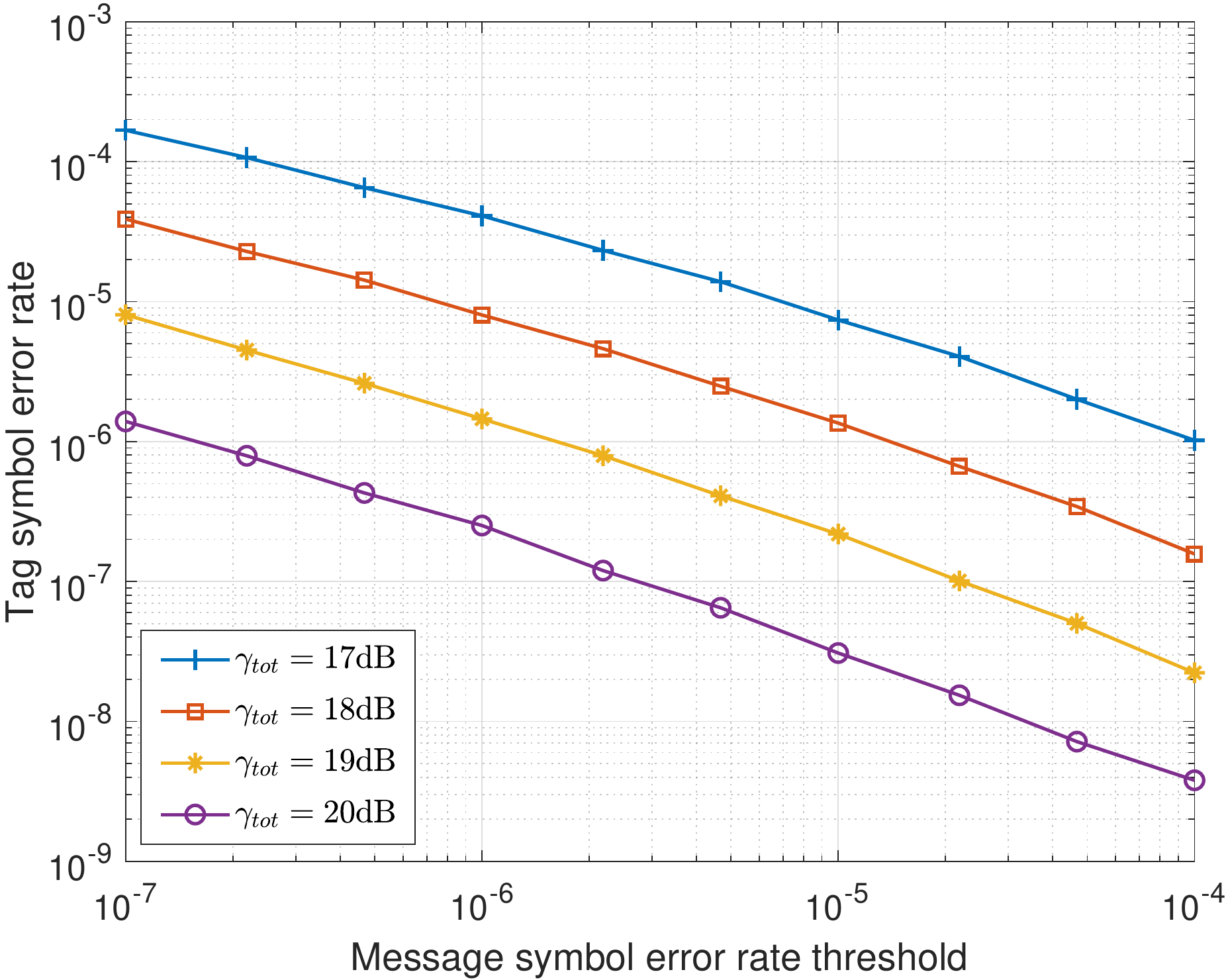}
	\caption{The trade-off between the message and tag error performance.}
	\label{TradeoffCurve}
\end{figure}

Notice that the purpose of Fig.~\ref{Message-basedPerformance} is to verify the correctness of the theoretical analysis, and the optimization problem in \eqref{generaloptimal} is not considered in the simulation. Now to understand the entire system performance, we show the solution to the problem in \eqref{generaloptimal} in Fig.~\ref{TradeoffCurve}. In this simulation, the system parameters are kept the same as in Fig.~\ref{NewUniformPerformance}, i.e., $N=128,\ L_m=4,\ L_t=2$. Through different $\gamma_{tot}$ and $\delta$ parameters, the trade-off between message and tag error performance is characterized in Fig.~\ref{TradeoffCurve}. We can observe from the trade-off curves that the tag SER decreases as the message SER requirement threshold increases. Both the tag and message SER can be reduced when the total system average power increases. The trade-off curves present the optimal tag SER performance under different system requirements for message SER, which can provide useful insights for a practical PLA system design. Reasonable operating points are dependent on the specific requirements of the IIoT system, which can be determined through the trade-off curves in Fig.~\ref{TradeoffCurve}. According to the reliability requirement and the power constraint of the system, we can obtain the expected authentication accuracy. As such, the system can operate under this certain transmitter power with the specific reliability and authentication accuracy. 

\begin{figure}[htbp]
	\centering
	\includegraphics[width=0.45\textwidth]{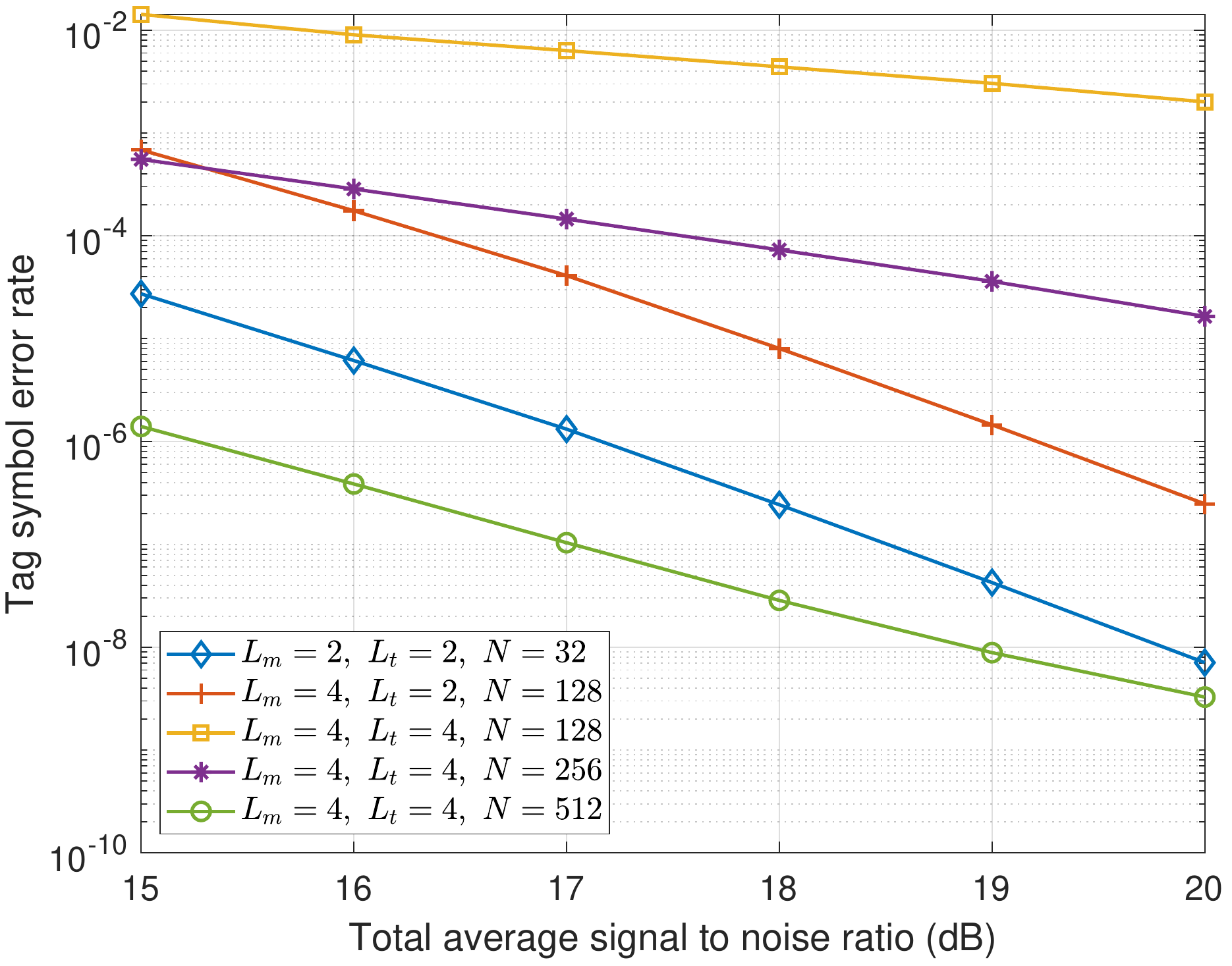}
	\caption{Tag SER performance under different system parameters.}
	\label{ModulationOrderandNumberAntennasImpacts}
\end{figure}

We then consider the impact of modulation orders and the number of receiving antennas on system performance. The tag SER performance under different parameters $L_m$, $L_t$ and $N$ is illustrated in Fig.~\ref{ModulationOrderandNumberAntennasImpacts}. In this simulation, a typical message SER threshold $\delta$ is set to $10^{-6}$. As shown in Fig.~\ref{ModulationOrderandNumberAntennasImpacts}, relatively small message and tag modulation order (e.g., $L_m=2$, $L_t=2$) can achieve a low tag SER even with a moderate number of receiving antennas (e.g., $N=32$). With other parameters unchanged, the tag SER increases dramatically when $L_t$ changes from $2$ to $4$. However, the tag SER can be reduced significantly by increasing the number of receiving antennas. Therefore, the system overall performance can be enhanced further by increasing the number of receiving antennas.
\section{Conclusions}\label{ConclusionSec}
In this paper, we proposed an active PLA mechanism, message-based tag embedding, for non-coherent massive SIMO-based IIoT systems. This paper is the first to show that active PLA can be realized without the need of pilot signal and channel estimation. We designed the optimal tag embedding scheme when the message constellation is given. Then we solved the power allocation problem to obtain the optimal tag SER performance. The trade-off curves between tag SER and message SER were depicted to offer a comprehensive understanding of the system performance. From the simulation results, we can conclude that message-based tag embedding is essential for the considered non-coherent system. The proposed authentication method can meet the specific power and error rate requirements of the IIoT system. Moreover, increasing number of receiving antennas can boost system reliability by reducing the impact of fading. The proposed PLA method might encounter a potential weakness in practical implementation. When jamming attacks occur, the receiver may not be able to accurately recover the message and the MAC. As a future work, we will consider the impacts of jamming attacks on our PLA scheme, and extend the application of our scheme to industrial scenarios with severe interference.

\appendices
\section{Proof of an Upper Bound of the Message SER}\label{ProofUpperBoundSec}
Notice that the average message SER of $x_1$ is less than for transmitting $\sqrt{|{m_1}|^2+|t_{1,Lt}|^2}$, because near constellation points produce a large message SER. Then the upper bound of $P_{em,1}$ can be derived as follows
\vspace{-1em}

\begin{small}
\begin{equation}
\begin{aligned}
P_{em,1}&=\frac{1}{L_t}\Bigg[{\rm P}\left(\hat{m}=m_2,\cdots,m_{L_m}|x_1=\sqrt{|{m_1}|^2+|t_{1,1}|^2}\right) \cr 
&\ \ \  +{\rm P}\left(\hat{m}=m_2,\cdots,m_{L_m}|x_1=\sqrt{|{m_1}|^2+|t_{1,2}|^2}\right)+\cdots\cr
&\ \ \  +{\rm P}\left(\hat{m}=m_2,\cdots,m_{L_m}|x_1=\sqrt{|{m_1}|^2+|t_{1,L_t}|^2}\right)\Bigg] \cr
&<{\rm P}\left(\hat{m}=m_2,\cdots,m_{L_m}|x_1=\sqrt{|{m_1}|^2+|t_{1,L_t}|^2}\right) \cr
&={\rm P}\left(\frac{||y||^2}{N}>B_1\right) \cr
&=1-G\left(\frac{NB_1}{A_{1,L_t}}\right).
\end{aligned}
\end{equation}
\end{small}Similarly, the average message SER of $x_{L_m}$ is less than transmitting $\sqrt{|{m_{L_m}}|^2+|t_{L_m,1}|^2}$, so the upper bound of $P_{em,L_m}$ can be derived as follows
\vspace{-1em}

\begin{small}
\begin{equation}
\begin{aligned}
P_{em,L_m}&<{\rm P}\left(\hat{m}\!=\!m_1,\cdots,m_{L_m-1}|x_{L_m}\!=\!\sqrt{|{m_{L_m}}|^2+|t_{L_m,1}|^2}\right) \cr
               &={\rm P}\left(\frac{||y||^2}{N}<B_{L_m-1}\right) \cr
               &=G\left(\frac{NB_{L_m-1}}{A_{L_m,1}}\right).
\end{aligned}
\end{equation}     
\end{small}Further, the average message SER of $x_i( i=2, \cdots, L_m-1)$ also has an upper bound due to the same reason above,
\vspace{-1em}

\begin{small}
\begin{equation}
\begin{aligned}
P_{em,i}&<{\rm P}\left(\hat{m}=m_1,\cdots,m_{i-1}|x_i=\sqrt{|{m_i}|^2+|t_{i,1}|^2}\right) \cr 
             &\ \ \ +{\rm P}\left(\hat{m}=m_{i+1},\cdots,m_{L_m}|x_i=\sqrt{|{m_i}|^2+|t_{i,L_t}|^2}\right)\cr
             &=1-G\left(\frac{NB_i}{A_{i,L_t}}\right)+G\left(\frac{NB_{i-1}}{A_{i,1}}\right).
\end{aligned}
\end{equation}
\end{small}From the above, an upper bound of $P_{em}$ can be derived as
\begin{equation}
\begin{aligned}
P_{em} &=\frac{1}{L_m}\sum \limits_{i=1}^{L_m}P_{em,i} \cr
             & < \frac{1}{L_m}\sum \limits_{i=1}^{L_m-1}\left[1-G\left(\frac{NB_i}{A_{i,L_t}}\right)+G\left(\frac{NB_i}{A_{i+1,1}}\right)\right] \cr 
             & =\frac{1}{L_m}\sum \limits_{i=1}^{L_m-1}\left\{1-G\left[Ng(r_i)\right]+G\left[Nh(r_i)\right]\right\},
\end{aligned}
\end{equation}  
where $g(r_i)=\frac{{R}\ln{\frac{{R}}{{r_i}^{L_t-1}}}}{{R}-{r_i}^{L_t-1}}$, $h(r_i)=\frac{{r_i}^{L_t-1}\ln{\frac{{R}}{{r_i}^{L_t-1}}}}{{R}-{r_i}^{L_t-1}}$. 

Therefore, an upper bound of message SER, $P_{em}^{u}$, can be represented by
\begin{equation}
P_{em}^u=\frac{1}{L_m}\sum \limits_{i=1}^{L_m-1}\left\{1-G\left[Ng(r_i)\right]+G\left[Nh(r_i)\right]\right\}.
\end{equation}

\section{Proof of the convex optimization problem}\label{ProofConvexSec}
Let $F(k)=1+G\left[Nu(e^{k})\right]-G\left[Nv(e^{k})\right]$ and $W(k)=1-G\left[Ng(e^{k})\right]+G\left[Nh(e^{k})\right]$. 

First, we prove that the objective function in \eqref{convex1} is a convex function for $0<k<\frac{\ln{R}}{L_t-1}$. $F(k)$ is a convex function for $0<k<\frac{\ln{R}}{L_t-1}$ according to the \textit{Lemma 2} in \cite{a9}. The objective function in \eqref{convex1} is a sum of $L_m$ convex functions $F(k_i)\ ( i=1,\cdots,L_m)$ and its Hessian matrix is a diagonal matrix, thus the Hessian matrix is positive definite and the objective function in \eqref{convex1} is a convex function for $0<k<\frac{\ln{R}}{L_t-1}$. 

Then, we prove that the left side of \eqref{convex2} is a convex function for $0<k<\frac{\ln{R}}{L_t-1}$. Note that $e^k-1$ is a basic convex function. The left side of \eqref{convex2} is a sum of $L_m$ convex functions $e^{k_i}-1\ (i=1,\cdots,L_m)$ and its Hessian matrix is a diagonal matrix, so it is also a convex function for $0<k<\frac{\ln{R}}{L_t-1}$. 

Finally, we prove that the left side of \eqref{convex3} is a convex function for $0<k<\frac{\ln{R}}{L_t-1}$. Let $W_1(k)=G\left[Ng(e^{k})\right]$ and $W_2(k)=G\left[Nh(e^{k})\right]$. Then we have $W'(k)=W'_2(k)-W'_1(k)$. The derivative of $G(z)$ is $f_Z(z)=\frac{1}{(N-1)!}z^{N-1}e^{-z}$. Then we have
\begin{equation}
\begin{aligned}
W'_1(k)&=f_Z\left[Ng(e^k)\right]Ng'(e^k) \cr
            &=\frac{1}{(N-1)!}\left[Ng(e^k)\right]^{N-1}e^{-Ng(e^k)}Ng'(e^k).
\end{aligned}
\end{equation}
Since $h(e^k)=g(e^k)\frac{e^{k(L_t-1)}}{R}=g(e^k)+\left[k(L_t-1)-\ln{R}\right]$, we can obtain that $h'(e^k)=\frac{1}{R}e^{k(L_t-1)}\left[g(e^k)(L_t-1)+g'(e^k)\right]$. Then we have
\begin{equation}
\begin{aligned}
W'_2(k)&=f_Z\left[Nh(e^k)\right]Nh'(e^k) \cr
            &=\frac{1}{(N-1)!}\left[Nh(e^k)\right]^{N-1}\cdot e^{-Nh(e^k)}Nh'(e^k) \cr
            &=\frac{1}{(N-1)!}\left[Ng(e^k)\frac{e^{k(L_t-1)}}{R}\right]^{N-1} \cr & \ \ \ \cdot e^{-N\left[g(e^k)+k(L_t-1)-\ln{R}\right]} \cr &\ \ \  \cdot N\left\{\frac{1}{R}e^{k(L_t-1)}\left[g(e^k)(L_t-1)+g'(e^k)\right]\right\} \cr
            &=W'_1(k)+\frac{1}{(N-1)!}N^Ng(e^k)^{N}e^{-Ng(e^k)}(L_t-1).
\end{aligned}
\end{equation}
Therefore, we can simplify the $W'(k)$ as follows
\begin{equation}
W'(k)=\frac{1}{(N-1)!}N^N(L_t-1)g(e^k)^{N}e^{-Ng(e^k)}.
\end{equation}
The second-order derivative of $W(k)$ can be obtained by

\begin{small}
\begin{equation}
W''(k)=\frac{N^{N+1}(L_t-1)g(e^k)^{N-1}e^{-Ng(e^k)}g'(e^k)\left[1-g(e^k)\right]}{(N-1)!},
\end{equation}
\end{small}where 

\begin{small}
\begin{equation}
g'(e^k)\!=\!\frac{(L_t\!-\!1)R\left\{e^{k(L_t-1)}\!-\!R\!+\!e^{k(L_t-1)}\left[\ln{R}\!-\!k(L_t\!-\!1)\right]\right\}}{\left[R-e^{k(L_t-1)}\right]^2}.
\end{equation}
\end{small}Let $q=e^{k(L_t-1)}$, then
\begin{equation}
g'(e^k)=\frac{(L_t-1)Rq}{(R-q)^2}(1+\ln{\frac{R}{q}}-\frac{R}{q}).
\end{equation}
Since $\ln{x}-x$ is a monotonically decreasing function for $x>1$ and $\frac{R}{q}>1$, then we have $1+\ln{\frac{R}{r}}-\frac{R}{r}<0$. Therefore, $g'(e^k)<0$ for $0<k<\frac{\ln{R}}{L_t-1}$. Then we know that $g(e^k)$ is a decreasing function when $0<k<\frac{\ln{R}}{L_t-1}$. Moreover,
\begin{equation}
    \lim_{k \to \frac{\ln{R}}{L_t-1}}g(e^k)=\lim_{q \to R}\frac{R\ln{\frac{R}{q}}}{R-q}=1
\end{equation}then $1-g(e^k)<0$ for $0<k<\frac{\ln{R}}{L_t-1}$. We can conclude that $W''(k)>0\ (0<k<\frac{\ln{R}}{L_t-1})$, thus $W(k)$ is a convex function for $0<k<\frac{\ln{R}}{L_t-1}$. The left side of \eqref{convex3} is a sum of $L_m-1$ functions $W(k_i)\ ( i=1,\cdots,L_m-1)$ and its Hessian matrix is a diagonal matrix. Therefore, the left side of \eqref{convex3} is a convex function for $0<k<\frac{\ln{R}}{L_t-1}$.

Above all, the optimization problem \eqref{convex} is a convex problem for $0<k<\frac{\ln{R}}{L_t-1}$.



\ifCLASSOPTIONcaptionsoff
  \newpage
\fi

\end{document}